\documentclass[10pt,twocolumn,letterpaper]{article}
\usepackage{cvpr}
\usepackage{times}
\usepackage{epsfig}
\usepackage{graphicx}
\usepackage{amsmath}
\usepackage{amssymb}
\usepackage{hyperref}
\usepackage{booktabs}
\usepackage{balance}
\usepackage{multirow}
\usepackage{url}
\usepackage{dsfont} % Add this line to include the bbm package
\usepackage[utf8]{inputenc}
\hypersetup{
    colorlinks=true,    % Enable colored links
    linkcolor=black,    % Color of internal links (e.g., references)
    citecolor=black,     % Color of citation links
    urlcolor=black       % Color of external links
}

\newif\ifijcbfinal
\ijcbfinaltrue

\ifijcbfinal
  \newcommand{\ijcbfinalcopy}{}
\fi

\DeclareUnicodeCharacter{2013}{--} % en dash
\DeclareUnicodeCharacter{200B}{} % zero-width space

\providecommand{\keywords}[1]
{
  \small	
  \textbf{\textit{Keywords---}} #1
}
\begin{document}
\ijcbfinalcopy

%%%%%%%%% TITLE
\title{Contextual Cross-Modal Attention for Audio-Visual Deepfake Detection and Localization}

\author{Vinaya Sree Katamneni and Ajita Rattani\\
University of North Texas at Denton\\
Denton, Texas, USA\\
{\tt\small vinayasreekatamneni@my.unt.edu; \tt\small ajita.rattani@unt.edu}}
%\and
%Ajita Rattani\\
%University of North Texas at Denton\\
%Denton, Texas\\
%{\tt\small ajita.rattani@unt.edu}
%}

\maketitle
\thispagestyle{empty}

%%%%%%%%% ABSTRACT
\begin{abstract}
In the digital age, the emergence of deepfakes and synthetic media presents a significant threat to societal and political integrity. 
Deepfakes based on multi-modal manipulation, such as audio-visual, are more realistic and pose a greater threat.   
Current multi-modal deepfake detectors are often based on the attention-based fusion of heterogeneous data streams from multiple modalities. However, the heterogeneous nature of the data (such as audio and visual signals) creates a distributional modality gap and poses a significant challenge in effective fusion and hence multi-modal deepfake detection.
In this paper, we propose a novel multi-modal attention framework based on recurrent neural networks (RNNs) that leverages contextual
information for audio-visual deepfake detection. The proposed approach applies
attention to multi-modal multi-sequence representations and learns the contributing features among them for deepfake detection and localization.
Thorough experimental validations on audio-visual deepfake datasets, namely FakeAVCeleb, AV-Deepfake1M, TVIL, and LAV-DF datasets, demonstrate the efficacy of our approach. Cross-comparison with the published studies demonstrates superior performance of our approach with an improved accuracy and precision by $3.47\%$ and $2.05\%$ in deepfake detection and localization, respectively. Thus, obtaining state-of-the-art performance. To facilitate reproducibility, the code and the datasets information is available at~\url{https://github.com/vcbsl/audio-visual-deepfake/}.

\end{abstract}


\section{Introduction}
\label{introduction}
With advances in deep-generative models~\cite{10.1145/3625547}, synthetic audio and visual media have become so realistic that they are often indistinguishable from authentic content for human eyes. However, synthetic media generation techniques used by malicious users to deceive pose a serious social and political threat~\cite{Hwang2020, citron, Tolosana2020deepfakesAB, Nguyen2019DeepLF, Li_2019_CVPR_Workshops,
8630761, 9010912,9157215}. 

In this context, visual (facial) deepfakes are generated using facial forgery techniques that depict human subjects with altered identities (i.e., face swapping), malicious actions (such as expression swapping), and facial attribute manipulation (such as skin color, gender, and age)~\cite{Choi_2018_CVPR, Nirkin_2019_ICCV, Xu_2018_CVPR}. Voice deepfakes, like facial deepfake technology, rely on advanced generative neural networks to synthesize audio that mimics the voice of a target speaker. Among them, Text-to-speech (TTS) voice deepfakes involve generating synthetic speech from text input that mimics a specific target speaker's voice~\cite{salvi2023timit}. Voice conversion-based deepfakes involve altering a person's voice to sound like another person while retaining the original content and linguistic style~\cite{sun2023ai}. 

Audio and visual deepfakes have been employed to attack authentication systems, impersonate celebrities and politicians, and defraud finance. As a countermeasure, several unimodal audio and visual deepfake detectors have been proposed~\cite{9157215,9578910,9360904, journals/corr/abs-1003-4083,chintha2020recurrent,hamza2022deepfake,pianese2022deepfake}. Lately, multi-modal deepfakes that manipulate multiple modalities, such as audio-visual, to create highly convincing and immersive fake content have shown staggering growth with advanced multimedia processing and generative AI capabilities~\cite{zhang2022deepfake}. These advanced multi-modal deepfake techniques leverage the strengths of different modalities to generate more realistic and impactful results.

\begin{figure*}
\centering
    \includegraphics[width=1.0\linewidth]{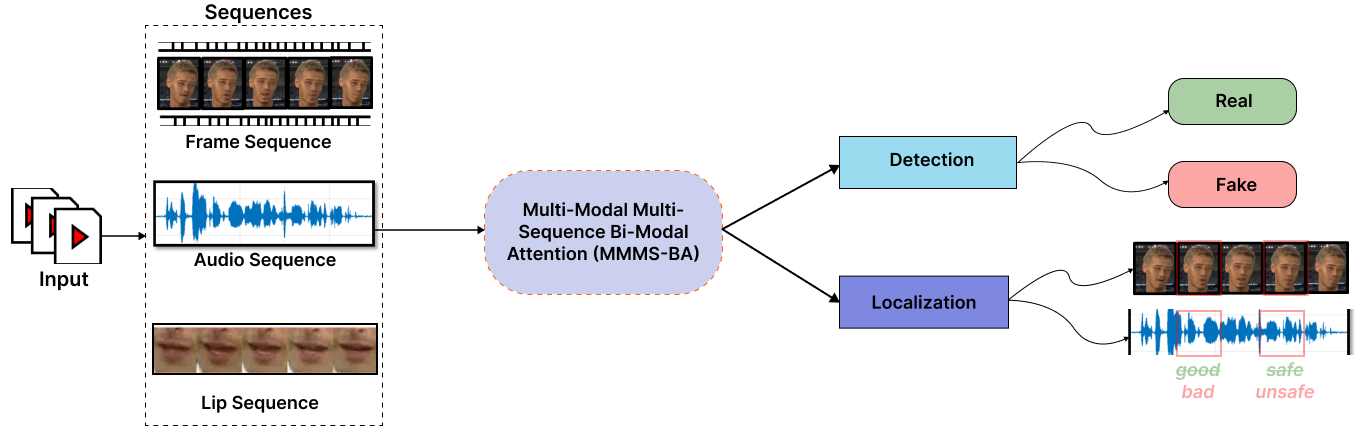}
    %\caption{This study proposes a contextual cross-attention-based framework for Deepfake Detection and Localization}
    \caption{Overview of our proposed audio-visual deepfake detection and localization framework. The audio-visual sequences extracted from the input video are processed using our proposed MMMS-BA approach for deepfake detection and localization.}
    \label{fig:img2b}
\end{figure*}

%Deepfakes evolution raises new challenges for detection methods, making the fight against deepfakes even more complex. A single detector may not always have access to the range or type of manipulated modalities beforehand. 
Existing unimodal deepfake detectors are primarily designed to detect a single type of manipulation, such as visual, acoustic, and text. 
Consequently, multi-modal deepfake detectors are being investigated to detect and localize multi-modal manipulations, collectively. Within the scope of this work, several audio-visual deepfake detection and localization techniques have been proposed%~\cite{9980296,10.1145/3476099.3484315, DBLP:journals/corr/abs-2108-05080, 9859289, poria-etal-2017-context, khalid2021fakeavceleb, 8840977,cai2023av, cai2023glitch, cai2022you, chen2023npvforensics, 10.1145/3394171.3413570, 
%~\cite{10095247, cheng2022voice, 9980296, Cozzolino2022AudioVisualPD, 10.1145/3476099.3484315, 10.1145/3394171.3413700, 
~\cite{raza2023multimodaltrace, feng2023self, ILYAS2023110124, 9350195, katamneni_nadimpalli_rattani_2023, sree2023mis,tian2023unsupervised, cozzolino2023audio, liu2023mcl, yu2023pvass, elpeltagy2023novel}. 
Deepfake detection aims at binary classification into pristine or deepfake. Localization aims to locate the start and end timestamps of manipulated audio-visual segments, thus facilitating a better understanding of deepfake detection results. Existing audio-visual deepfake detectors are often based on the fusion of heterogeneous streams using feature concatenation and employing attention mechanism~\cite{zhou2021joint, 9350195, zhao2023fine, masood2023attention,wang2022audio,asha2024defensive, oorloff2024avff}. 
Deepfake localization approaches are either anchor-based~\cite{gao2018ctap, gao2017turn} that utilize a sliding window approach to detect deepfake segments or boundary prediction based~\cite{bagchi2021hear,cai2022you,cai2023glitch,zhang2023ummaformer} that
focuses on predicting the boundary of fake segments in a video using boundary matching loss.

The major \textbf{limitations} of the current attention-based fusion approaches for audio-visual deepfake detection stem from the heterogeneous nature of audio and visual signals. Consequently, the unique capabilities of each modality (modality-specific features) are not utilized effectively in the fusion process. Furthermore, noise in one modality adversely affects the overall performance of the multi-modal framework. %This limits the ability of existing approaches to fully leverage modality-specific features for audio-visual deepfake detection. 
These methods also fail to explicitly model the interactions between different modalities, such as the relationship between audio, full face, and lip movement sequences. As a result, correlation and dependencies between modalities are not fully captured, leading to suboptimal performance in deepfake detection. 
Moreover, current approaches do not incorporate contextual information which refers to consistent and meaningful patterns across sequences, both within and across modalities.
% Lastly, most current fusion-based approaches focus solely on audio-visual detection without integrated mechanisms for localization. Given that deepfakes have become more content-driven generating partially fake media~\cite{cai2023glitch,cai2023av,zhang2023ummaformer}, integrated deepfake detection as well as localization is crucial. 
Lastly, most current fusion-based approaches focus solely on audio-visual detection without integrated mechanisms for localization. Given that deepfakes have become more content-driven generating partially fake media~\cite{cai2023glitch,cai2023av,zhang2023ummaformer}, integrated deepfake detection as well as localization is crucial.

This work \textbf{proposes} a recurrent neural network-based multi-modal multi-sequence attention framework for audio-visual deepfake detection, called Multi-Modal Multi-Sequence Bi-Modal Attention (MMMS-BA). Our framework focuses on relevant features across modality pairs, leveraging attention from neighboring sequences and multi-modal representations for enhanced representation learning. %The primary challenge in audio-visual deepfake detectors is the effective utilization of information from heterogeneous modalities. 
%To address this, our novel method emphasizes inter-modality relations computed between sequences and their context. By utilizing information from multiple modalities, the method considers not only the relation between two modalities within the same sequence but also their relatedness with modalities across surrounding sequences. This is achieved through multi-sequence attention instead of single-sequence attention.
Our proposed MMMS-BA performs deepfake detection as well as localization using classification and regression heads. %It also accounts for missing modalities through masking layers. Masking layers indicate missing or irrelevant segments of the input, allowing the model to ignore those missing segments during training and inference. %The padding ensures uniform input length by adding placeholder values, enabling efficient processing. 
Figure~\ref{fig:img2b} illustrates our proposed MMMS-BA framework for audio-visual deepfake detection and localization.

In summary, the \textbf{contributions} of this work are as follows: 
\begin{enumerate}

\item A novel approach for audio-visual deepfake detection and localization (MMMS-BA) based on contextual cross-attention mechanism utilizing multi-modal multi-sequence information.

%\item An effective attention framework (MMMS-BA) that enables the model to apply attention to the sequences and/or multi-modal representations.

\item Extensive evaluation on the publicly available audio-visual deepfake detection and localization datasets namely, AV-DeepFake1M~\cite{cai2023av}, FakeAVCeleb~\cite{DBLP:journals/corr/abs-2108-05080}, LAV-DF~\cite{cai2022you}, and TVIL~\cite{zhang2023ummaformer}.

\item Cross-comparison of MMMS-BA with the published work on audio-visual multi-modal deepfake detection and localization.

\end{enumerate}
% Our contributions to the field of deepfake detection are as follows:

% \begin{enumerate}
%     \item We introduce a novel contextual cross-attention mechanism that dynamically facilitates the integration of multi-modal data. 
%     \item We propose a framework that adapts to the unique challenges posed by various modalities and combinations of them. 
%     \item We propose a module to localize and ground the deepfakes detected.
%     \item Extensive evaluation on AV-DeepFake-1M~\cite{cai2023av}, FakeAVCeleb~\cite{DBLP:journals/corr/abs-2108-05080}, LAV-DF~\cite{cai2022you}, and TVIL~\cite{zhang2023ummaformer} datasets to validate the superiority of our approach.

% \end{enumerate}

This paper is summarized as follows: Section~\ref{relatedwork} discusses related work on audio-visual deepfake detection and localization. The proposed approach is detailed in Section~\ref{approach}. Section~\ref{datasets} covers the datasets, evaluation metrics, and results. An ablation study analyzing varied modalities and attention mechanisms is presented in Section~\ref{ablation}. Conclusion and future research directions are discussed in Section~\ref{conclusion}.

\section{Related Work}
\label{relatedwork}
%The field of deepfake detection has witnessed considerable advancements, particularly in the realm of multi-modal manipulation detection. This section synthesizes the contributions, including the development of sophisticated datasets and innovative detection methodologies required to tackle the complexities that are inherited by deepfake generation techniques.
In this section, we discuss the work related to audio-visual multi-modal deepfake detection and localization.

\subsection{Audio-Visual Deepfake Detection}

Audio-visual deepfake detection techniques employ audio and visual signals to detect multi-modal manipulation. A foundational work in this area,~\cite{10.1145/3476099.3484315} assembled the FakeAVCeleb dataset consisting of audio and visual deepfakes and benchmarked various audio-visual deepfake detectors based on ensemble-based voting scheme and multi-modal convolutional neural network (CNN) based on feature concatenation. 

In particular, most of the existing audio-visual deepfake detectors are based on attention mechanism-based feature concatenation~\cite{zhou2021joint,zhao2023fine, 9350195, shao2024detecting, masood2023attention, wang2022audio, asha2024defensive, oorloff2024avff} to learn informative multi-modal features for deepfake detection. %The approaches as introduced employ this trend by aligning and integrating audio and visual features. Fused features are then processed using an attention mechanism.
Studies in~\cite{10.1145/3394171.3413570, nekadi2020siamese} used a siamese network architecture for emotion recognition from audio and visual cues incorporating contrastive loss. Deepfake videos are detected by analyzing discrepancies in emotional cues between audio and visual modalities.
%used Siamese Network which uses uni-modal features fed into an emotion recognition system to compare disagreement in emotional cues from both modalities in a video for deepfake detection. 
Studies in~\cite{10.1145/3394171.3413700,10095247, liu2023magnifying, cheng2023voice, zhang2021deepfake, lewis2020deepfake, liu2023mcl} explicitly model the disagreement between the embeddings of the multiple modalities using contrastive loss for deepfake detection.

% This approach allows for a more dynamic examination of the temporal relationships between the two modalities, enhancing the detection capabilities of the system.
%Similarly, additional research by various scholars, including those in~\cite{9350195, shao2024detecting, masood2023attention, wang2022audio, asha2024defensive, oorloff2024avff}, has embraced soft attention within their modality fusion processes. These studies utilize attention blocks to refine the fused audio-visual representations, thereby generating more precise and relevant feature representations for subsequent analysis steps. However, these techniques often do not fully address performance challenges in scenarios where one modality is manipulated or missing, focusing instead on the integrated features without isolating the distinct characteristics inherent to each modality.

%to single-modality embeddings to highlight disparities in audio-visual pairs. 

%~\cite{} uses contrastive loss to assess the consistency between voice and face. 

%The challenge of detecting inconsistencies in phoneme-viseme mappings within deepfake videos has been extensively studied in recent years. 
The work in~\cite{agarwal2020detecting, agarwal2023watch} and \cite{chen2023npvforensics} explores the mismatch between phonemes (distinct units of sound in speech) and visemes (the visual representation of phonemes) for deepfake detection. These studies specifically focus on inconsistencies in the lip region concerning the audio which deepfake generation approaches often struggle to replicate incorrectly, thus offering a potential for deepfake detection~\cite{agarwal2020detecting, agarwal2023watch}.

\begin{figure*}
\centering
    \includegraphics[width=1.00\linewidth]{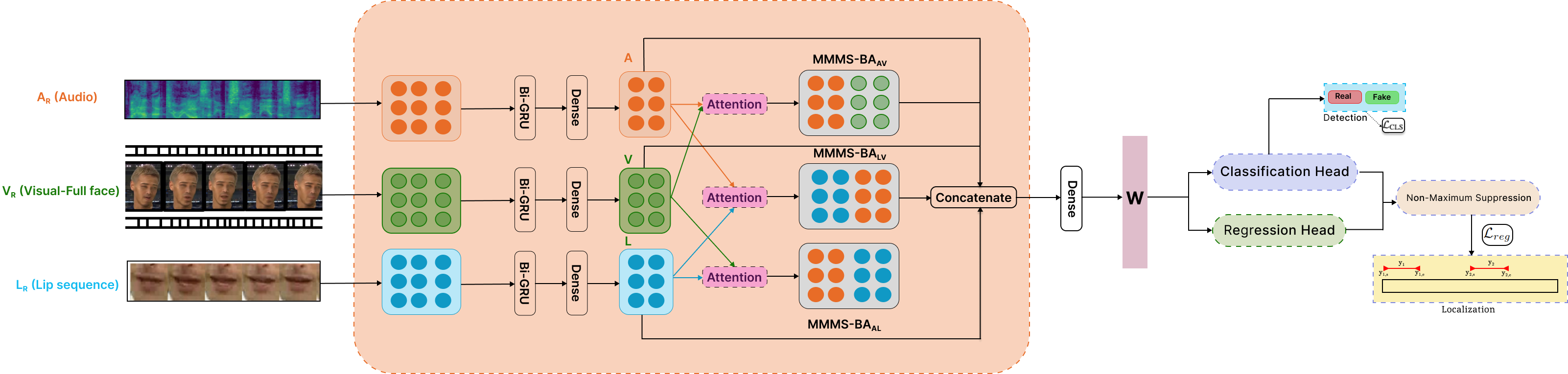}
    % \caption{Multi-Modal Multi Sequence Bi-Modal Attention-based Deepfake Detector (MMMS-BA Framework)}
    \caption{Illustration of the proposed Multi-Modal Multi-Sequence Bi-modal Attention (MMMS-BA) model for audio-visual deepfake detection and localization.}
    \label{fig:img2a}
\end{figure*}

% Concurrently, the introduction of the AV-Deepfake1M~\cite{cai2023av} dataset amplifies a pioneering effort to emulate and check the generation of content with embedded audio-visual manipulations across more than 2,000 subjects in over a million videos. This dataset's inception is pivotal, with a decrease in detection performance by state-of-the-art methodologies against such complex manipulations. Addressed the need for datasets that can challenge and refine the detection capabilities of existing models.

% The Boundary Aware Temporal Forgery Detection (BA-TFD)~\cite{cai2023glitch} emerges as a novel baseline method employing a 3D Convolution Neural Network architecture adept at capturing multi-modal manipulations. The subsequent enhancement, BA-TFD+, through the integration of a Multi-scale Vision Transformer and a training regime incorporating contrasting, frame classification, boundary matching, and multi-modal boundary matching loss functions, showcases superior performance in both the detection and temporal localization of deepfake content across benchmark datasets, including the newly proposed LAV-DF. These advancements not only signify a leap in deepfake detection methodologies but also highlight the complexity and necessity of temporal forgery localization within the broader spectrum of deepfake challenges.

\subsection{Localization of Deepfakes}
Localization aims to ground the time intervals in an input where manipulation has occurred. %Since the task of deepfake localization is similar to the temporal action localization work in ~\cite{cai2023av,cai2023glitch,cai2022you,zhang2023ummaformer} adopted it. 
Audio-visual deepfake localization-based approaches can be categorized into two: anchor-based~\cite{gao2018ctap, gao2017turn} and those based on the prediction of boundaries of fake segments in a video~\cite{bagchi2021hear,cai2022you, cai2023glitch, zhang2023ummaformer,cai2023av}. Anchor-based approaches make use of a sliding window over the input to ground the manipulated segments. In contrast, prediction-based approaches process the entire video and determine the fake segments using the proposed boundary-sensitive network~\cite{bagchi2021hear}. 
% global information here is going through the entire video and determining the segments of fake and then localizing by predicting the start and end time stamps of the fake segments identified. 
%BoundarySensitive Network (BSN), which adopts “local to global” fashion to locally combine
% high probability boundaries as proposals and globally retrieve candidate proposals using proposal-level feature as shown in Fig 1. In detail, BSN generates proposals in three
% steps. First, BSN evaluates the probabilities of each temporal location in video whether
% it is inside or outside, at or not at the boundaries of ground truth action instances, to
% generate starting, ending and actionness probabilities sequences as local information.
% Second, BSN generates proposals via directly combining temporal locations with high
% starting and ending probabilities separately. Using this bottom-up fashion, BSN can
% generate proposals with flexible durations and precise boundaries. Finally, using features composed by actionness scores within and around proposal, BSN retrieves proposals by evaluating the confidence of whether a proposal contains an action. These
% proposal-level features offer global information for better evaluation.
% Given a segment focuses on predicting the class labels (real or fake) and segment boundaries of all fake segments present in a given video.

The foundational study in~\cite{10.1145/3394171.3413700} introduced a novel method focused on detecting and localizing discrepancies between audio and visual modalities based on temporal mismatches and unnatural movements within videos. 
% The approach, described in their paper, utilizes a Modality Dissonance Score (MDS) to measure inconsistencies between the audio and visual streams. 
%This technique %leverages a deep learning network to 
%discern and localize temporal mismatches and unnatural movements within videos. 
%The framework established in this study set the groundwork for more advanced techniques.
Building on their initial work~\cite{cai2022you}, the authors introduced a new dataset, Localized Audio-Visual Deepfake (LAV-DF), along with the multi-modal deepfake detection that incorporates contrastive loss function to differentiate between real and fake segments by pushing apart the feature representations of genuine and forged segments in the feature space. In addition, the loss of boundary matching precisely predicts the boundaries (start and end points) of the forged segments.
%specifically designed to enhance the learning and evaluation of temporal forgery localization approaches. 
%The proposed multi-modal method
%, known as Boundary Aware Temporal Forgery Detection (BA-TFD), 
%incorporates contrastive, boundary matching, and frame classification loss functions. Contrastive loss is employed to differentiate between real and fake segments by pushing apart the feature representations of genuine and forged segments in the feature space. Boundary matching loss precisely predicts the boundaries (start and end points) of the forged segments. 
%Frame classification loss classifies the frame as real or fake. 
%This approach enhances the capability to accurately predict the boundaries of fake segments.
% thus improving the overall effectiveness of the localization technique.

Work in~\cite{cai2023glitch} advances the method in~\cite{cai2022you} further by introducing a framework (named BA-TFD+) that integrates a multi-scale vision transformer with a 3D CNN architecture and trained with contrastive, frame classification, boundary matching and multi-modal boundary matching loss functions for improved precision and reliability of deepfake detection and localization. %This method enhances the earlier approach by incorporating additional loss functions and a comprehensive boundary localization module for improved precision and reliability of deepfake detection and localization. 
% The BA-TFD+ system is thoroughly evaluated on the LAV-DF and other benchmark datasets, demonstrating its superiority in detecting and localizing content-driven audio-visual forgeries.

%In a parallel development,  
Another work~\cite{zhang2023ummaformer} in this direction introduces a framework named UMMAFormer, a universal, transformer-based framework adapted to audio and visual multi-modalities to detect temporal forgeries using anomaly detection based on self-attention mechanism. %Later authors in ~\cite{cai2023av} curated a novel audio-visual deepfake dataset to facilitate localization using content-driven manipulation and benchmarked the existing localization approaches~\cite{cai2022you,cai2023glitch,zhang2022actionformer,zhang2023ummaformer} on this dataset. 

\section{Methodology}
\label{approach}
\subsection{Problem Formulation}
Our proposed framework aims to detect and localize deepfakes by focusing on the relationship between audio, visual and lip movement sequences by harnessing the information embedded within each modality and across their intersection. 

%, drawing inspiration from the contextual inter-modal attention for multi-modal analysis in ~\cite{ghosal2018contextual}, 
 
For a given video $D$ that contains $N$ sequences, where each sequence $d_i$ is composed of $3$ different modalities (full visual face, lip region, and audio), the formulation can be represented as follows:
\begin{equation}
\label{eq1}
D = \{d_i\}_{i=1}^N  \quad ; \quad d_i = \{x_{i}^{v}, x_{i}^{l}, x_{i}^{a}\}
\end{equation}

Here, each sequence $d_i$ consists of three modalities represented by $x_{i}^{m}$ for $m \in \{v,l, a\}$ for the full visual face, lip region, and audio modality.

The classification head detects the modified sequences and the regression head is employed to localize fake segments. For the video $D$ with $N$ sequences, the entire video $D$ is classified as fake if at least one sequence is classified as fake. %For instance, let $N_f$ be the number of fake sequences detected. 

The localization of the segments is represented as \( Y = \{y_1, y_2, ..., y_{N_f}\} \) where $N_f$ are the number of fake segments. For a fake sequence $i$, \( y_i \) represents the \(i^{th}\) sequence output. Each instance \( y_i = (s_i, e_i) \) is defined by its starting time \( s_i \) and ending time \( e_i \) of the sequence. Figure~\ref{fig:img2a} illustrates the step-by-step process involved in our MMMS-BA which includes feature extraction and processing (as described below), and classification and regression heads for deepfake detection and localization. The following sub-sections provide details on each of these steps.

\subsection{Multi-modal Multi-sequence - Bi-modal Attention (MMMS-BA) Framework}
\label{3.2}
%To leverage multi-modal and contextual information for predicting the deepfakes. 
Sequences in a video represent the time series information and the classification of a sequence would have a relation with the other sequences. To model the relationship with the neighboring sequences and multiple modalities, we propose a recurrent neural network-based multi-modal attention framework named Multi-Modal Multi-Sequence Bi-modal Attention (MMMS-BA). Figure~\ref{fig:img2a} shows the steps involved in applying the attention mechanism to the input sequences.

The MMMS-BA framework processes and analyzes multi-modal data across sequences in an input video. Assuming a particular video has $N$ sequences, the raw sequence levels are represented as \(x_{i}^{v}\) for the full visual face, \(x_{i}^{a}\) for the audio and \(x_{i}^{l}\) for the lip sequence from equation ~\ref{eq1}.
% \subsubsection{Bi-GRU and Dense Layers Processing}
Three separate Bi-GRU layers with forward and backward state concatenation are first applied to the full visual face \(x_{i}^{v}\), audio \(x_{i}^{a}\), and lip sequence \(x_{i}^{l}\) representations followed by the fully connected dense layers, resulting in $L$ (lip region), $V$ (full visual face), and $A$ (audio) embeddings. Finally, pairwise attentions are computed on various bi-modal combinations of three modalities - $(V, L)$, $(L, A)$ \& $(A, V)$ as explained in the section~\ref{bmaa}. 

\subsubsection{Bi-modal Attention}
\label{bmaa}
Modality representations of $V$ \& $L$ are obtained from the Bi-GRU network and hence contain the contextual information of the sequences for each modality. Figure $1$ in the Appendix provides additional details on the computation of bi-modal attention for each modality pair. At first, we compute a pair of matching matrices $M1$ and $M2$ over two representations that account for the cross-modality information.
    
    \begin{equation}
    \label{eq2}
    M_1 = V \cdot L^T \quad \text{and} \quad M_2 = L \cdot V^T
    \end{equation}

\subsubsection{Multi-sequence Attention} 
As mentioned earlier, we aim to leverage the contextual information of each sequence for the prediction. The probability distribution scores ($K_1$ and $K_2$) are computed over each sequence of bi-modal attention matrices $M_1$ and $M_2$ (refer to equation ~\ref {eq2}) using a softmax function. This essentially computes the attention weights for the contextual sequences. Finally, soft attention is applied over the multi-modal multi-sequence attention matrices to compute the modality-wise attentive representations, i.e., $O_1$ and $O_2$ explained below.
    
\begin{equation}
\label{eq3}
    K_1(i, j) = \frac{e^{M1(i,j)}}{\sum_{k=1}^{N} e^{M1(i,k)}} \quad \text{for} \ i, j = 1,2, .., N
\end{equation}
\begin{equation}
\label{eq4}
    K_2(i, j) = \frac{e^{M2(i,j)}}{\sum_{k=1}^{N} e^{M2(i,k)}} \quad \text{for} \ i, j = 1,2, .., N
\end{equation}
    
\begin{equation}
\label{eq5}
    O_1 = K_1 \cdot L \quad \text{and} \quad O_2 = K_2 \cdot V
\end{equation}
where $M_1$ and $M_2$ are the bi-modal attention matrices between $V$ and $L$. $i$ and $j$ represent sequences in the input.

\subsubsection{Multiplicative Gating and Concatenation} 
A multiplicative gating function is computed between the multi-modal sequence-specific representations of each modality $O_1$ \& $O_2$ (refer equation~\ref{eq5}) and the corresponding bi-modal pair. This element-wise matrix multiplication assists in attending to the important components of multiple modalities and sequences. Attention matrices $A_1$ \& $A_2$ are then concatenated to obtain the pair-wise attention between $V$ and $L$.
    
\begin{equation}
\label{eq6}
    A_1 = O_1 \odot V \quad \text{and} \quad A_2 = O_2 \odot L
\end{equation}

\begin{equation}
\label{eq7}
    MMMS\text{-}BA_{VL} = \text{concat}[A_1, A_2]
\end{equation}

\noindent Similar to $MMMS\text{-}BA_{VL}$ in equation~\ref{eq7}, we follow the same procedure to compute $MMMS\text{-}BA_{AV}$, and $MMMS\text{-}BA_{AL}$. 
Finally, motivated by the residual skip connection network, the pairwise attention representations $MMMS\text{-}BA_{VL}$, $MMMS\text{-}BA_{AV}$, and $MMMS\text{-}BA_{AL}$ are concatenated with the individual modalities ($V$, $A$, and $L$) to increase gradient flow to the lower layers. This concatenated feature vector $W$ (see Figure~\ref{fig:img2a}) is then used for deepfake detection.

\subsection{Localization}
For the sequences classified as fake, their corresponding timestamp in the input video are localized as the fake segments. %The resulting fake segments for a sequence $i$ is (\(s_i, e_i\)). 
The localization of the segments is represented as \( Y = \{y_1, y_2, ..., y_{N_f}\} \) where $N_f$ are the number of fake segments. For a fake sequence $i$, \( y_i \) represents the \(i^{th}\) sequence output. Each instance \( y_i = (s_i, e_i) \) is defined by its starting time \( s_i \) and ending time \( e_i \) of the sequence. These segments are further processed using Non-Maximum Suppression (NMS) ~\cite{bodla2017soft} to remove highly overlapping instances, leading to the final localization timestamps.

% Each feature on the feature set \( W \) outputs a liveness score or probability \( p(l) \) and the corresponding temporal boundaries (\( s, e \)), which are then used to localize the deepfake.
\subsection{Loss Function} 
The model's overall learning process involves minimizing the combined loss as follows:
\begin{equation}
\label{eq8}
\mathcal{L} = \sum_{N}\left( \mathcal{L}_{cls} + \lambda_{reg}\mathds{1}_{c_i}\mathcal{L}_{reg} \right) / N_f,
\end{equation}

where \( N_f \) is the total number of fake sequences. \( \mathds{1}_{c_i} \) is an indicator function that denotes if a sequence \( i \) is fake with value equal to $1$ otherwise $0$. \( \mathcal{L} \) is applied and averaged in all sequences during training. \( \lambda_{reg} \) is a coefficient that balances classification and regression loss. We set \( \lambda_{reg}=1 \) by default.

Notably \( \mathcal{L}_{cls} \) uses focal loss~\cite{lin2017focal} to classify sequences as real or fake. \( \mathcal{L}_{reg} \) adopts a differentiable IoU loss from ~\cite{rezatofighi2019generalized}. \( \mathcal{L}_{reg} \) is only enabled when the current sequence is fake.

\section{Experimental Validations}
\label{datasets}
\subsection{Datasets}
Evaluation of the proposed method is conducted on publicly available audio-visual AV-DeepFake1M~\cite{cai2023av}, FakeAVCeleb~\cite{DBLP:journals/corr/abs-2108-05080}, LAV-DF~\cite{cai2022you}, and TVIL~\cite{zhang2023ummaformer} deepfake datasets. Table ~\ref{tab:datasets} provides details on the datasets used in this study. More details about the datasets are available in the Appendix, section $1$.

% \noindent \textbf{FakeAVCeleb:} The FakeAVCeleb dataset~\cite{DBLP:journals/corr/abs-2108-05080} is a collection of videos with audio and video manipulations of celebrities that have been generated using various deepfake techniques. 
% The dataset is created by selecting videos from the VoxCeleb2~\cite{chung18b_interspeech} dataset, featuring $500$ celebrities. 
%The videos in this dataset are clean, featuring only one person's frontal face without any occlusion. 
% The dataset is well-balanced and annotated in terms of gender, race, geography, and visual and audio manipulations, making it useful for training deep learning models that can generalize well on unseen test sets.  
% We chose this dataset for our experiments for multi-modal detection because it contains both audio and visual manipulations, as well as a variety of deepfake generation techniques.  
% We have used the gender and race-balanced version of the training and test set in this study.
% Adjusted Table for Cross-dataset Results
\begin{table*}[!ht]
\centering
\caption{Details on the datasets used in this study.}
\label{tab:datasets}
\scalebox{0.87}{
\begin{tabular}{|l|c|c|c|c|c|}
\hline
\textbf{Dataset} & \textbf{Year} & \textbf{Tasks} & \textbf{Manipulated Modality} & \textbf{Manipulation Method} & \textbf{\#Subjects} \\ \hline
FakeAVCeleb~\cite{khalid2021fakeavceleb} & 2021 & Detection & Audio and Visual & Re-enactment & 500 \\ \hline
LAV-DF~\cite{cai2022you} & 2022 & Detection and Localization & Audio and Visual & Content-driven & 153 \\ \hline
TVIL~\cite{zhang2023ummaformer} & 2023 & Detection and Localization & Audio and Visual & Inpainting forgery & N/A \\ \hline
AV-Deepfake1M~\cite{cai2023av} & 2023 & Detection and Localization & Audio and Visual & Content-driven & 2068 \\ \hline
\end{tabular}}
\end{table*}

\subsection{Implementation Details}
For the deepfake detection task using MMMS-BA, our model architecture includes bidirectional GRUs with $300$ neurons each, followed by a dense layer comprising $100$ neurons. This dense layer ensures that the input features from all three modalities, i.e., full visual face, lip sequences, and audios are projected to the same dimensions, facilitating cohesive integration of information. Dropout regularization is applied with a rate of $0.3$ across all layers, including the Bi-GRU layers. ReLU activation functions are utilized in the dense layers, while softmax activation is employed in the final classification layer, and ReLU (to ensure nonnegative values for start and end timestamps) is employed with Differential Intersection over Union (DIoU) loss, which measures the accuracy of the predicted timestamps against the ground truth.

The hyperparameters, including activation functions and dropout probabilities, were selected through a grid-search process using the validation set. The Adam optimizer with an exponential decay learning rate scheduler is used for optimization, and an early stopping strategy is implemented to determine the optimal training for each model. During training, we use a batch size of $32$. We conducted experiments using different combinations of bi-modal inputs, encompassing full visual face, lip sequences, and audio modalities. %Specifically, we explore all valid combinations where only one modality is taken at a time (uni-modal), any two modalities are taken at a time (bi-modal), and all three modalities are used simultaneously (tri-modal) as part of our ablation study.

\noindent \textbf{Evaluation Metrics}: For performance evaluation, we used standard evaluation metrics commonly used for deepfake detection, such as area under the ROC Curve~(AUC), partial AUC (pAUC) (at $10$\% False Positive Rate~(FPR)), and Equal Error Rate~(EER) similar to the studies on deepfake detection~\cite{Nadimpalli2022GBDFGB,khalid2021fakeavceleb}. 
%To evaluate fairness, we analyzed binary classification metrics, as deepfake detection is a binary classification task. 
% We followed the evaluation procedure established by the National Institute of Standards and Technology (NIST) and assessed the overall classification accuracy (ACC), along with the true positive rate~(TPR), and false-positive rate~(FPR) similar to the studies on the deepfake detectors~\cite{ Nadimpalli2022GBDFGB}.
For localization, we used average precision (AP) and average recall (AR) similar to the published work in~\cite{cai2023glitch,zhang2023ummaformer,cai2023av} at different thresholds to evaluate the model.

% \subsection{Results and Analysis}
%\label{analysis}
\subsection{Performance of Deepfake Detection}
Tables~\ref{table:cross-dataset-results-fakeavceleb}  and~\ref{table:cross-dataset-results-avdf1m} summarize the performance of our MMMS-BA model on intra- and cross-dataset evaluation when trained on FakeAVCeleb and AV-Deepfake1M datasets. 
For the AV-Deepfake1M dataset, training and validation sets are available, and the testing set has not yet been released. In this regard, we have used the training set for both training ($85$ \%) and validation ($15\%$). While the actual validation set is used for testing the trained models. 

% Adjusted Table for Intra-dataset Results
% \begin{table}[htbp]
% \centering
% \caption{Intra-dataset Evaluation Results}
% \label{table:intra-dataset-results}
% \scalebox{0.77}{
% \begin{tabular}{l|c|c|c|c|c|c}
% \hline
% \textbf{Dataset}            & \textbf{AUC} & \textbf{pAUC} & \textbf{EER} & \textbf{ACC} & \textbf{TPR} & \textbf{FPR} \\ \hline
% FakeAVCeleb                 & 0.990        & 0.977         & 0.029        & 0.979        & 0.965        & 0.039        \\ \hline
% AV-Deepfake1M               & 0.979        & 0.962         & 0.051        & 0.968        & 0.954        & 0.062        \\ \hline
% \end{tabular}}
% \end{table}
\begin{table}[h]
\centering
\caption{Intra- and cross-dataset evaluation of MMMS-BA when trained on FakeAVCeleb dataset.}
\label{table:cross-dataset-results-fakeavceleb}
\scalebox{0.87}{
\begin{tabular}{l|c|c|c|c|c|c}
\hline
\textbf{Testing Dataset} & \textbf{AUC} & \textbf{pAUC} & \textbf{EER} & \textbf{ACC} & \textbf{TPR} & \textbf{FPR} \\ \hline
FakeAVCeleb              & 0.989        & 0.977         & 0.029        & 0.979        & 0.965        & 0.039        \\ \hline
AV-Deepfake1M            & 0.909        & 0.884         & 0.178        & 0.821        & 0.893        & 0.249        \\ \hline
LAV-DF                   & 0.958        & 0.938         & 0.078        & 0.932        & 0.912        & 0.081        \\ \hline
TVIL                     & 0.942        & 0.920         & 0.126        & 0.947        & 0.927        & 0.093        \\ \hline
\end{tabular}}
\end{table}

% \begin{table}
% \centering
% \caption{Intra- and cross-dataset evaluation on \textbf{AV-Deepfake1M}.}
% \label{table:cross-dataset-results-avdf1m}
% \scalebox{0.85}{
% \begin{tabular}{l|c|c|c|c|c|c}
% \hline
% \textbf{Testing Dataset} & \textbf{AUC} & \textbf{pAUC} & \textbf{EER} & \textbf{ACC} & \textbf{TPR} & \textbf{FPR} \\ \hline
% % FakeAVCeleb & FakeAVCeleb  & 0.990        & 0.977         & 0.029        & 0.979        & 0.965        & 0.039        \\ \hline
% % FakeAVCeleb               & AV-Deepfake1M            & 0.909       & 0.884         & 0.178        & 0.821        & 0.893        & 0.249        \\ \hline
% % FakeAVCeleb               & LAV-DF                   & 0.958        & 0.938         & 0.078       & 0.932       & 0.912        & 0.081       \\ \hline
% % FakeAVCeleb               & TVIL                     & 0.942        & 0.920         & 0.126        & 0.947        & 0.927        & 0.093        \\ \hline

% AV-Deepfake1M               & 0.979        & 0.962         & 0.051        & 0.968        & 0.954        & 0.062        \\ \hline
% FakeAVCeleb              & 0.955        & 0.935         & 0.088        & 0.947        & 0.938        & 0.108        \\ \hline
% LAV-DF                   & 0.968        & 0.952         & 0.068        & 0.956        & 0.941        & 0.074        \\ \hline
% TVIL                     & 0.932        & 0.912         & 0.118        & 0.913        & 0.890        & 0.145        \\ \hline

% \end{tabular}}
% \end{table}

\begin{table}[ht]
\centering
\caption{Intra- and cross-dataset evaluation of MMMS-BA when trained on AV-Deepfake1M dataset.}
\label{table:cross-dataset-results-avdf1m}
\scalebox{0.87}{
\begin{tabular}{l|c|c|c|c|c|c}
\hline
\textbf{Testing Dataset} & \textbf{AUC} & \textbf{pAUC} & \textbf{EER} & \textbf{ACC} & \textbf{TPR} & \textbf{FPR} \\ \hline
AV-Deepfake1M             & 0.979        & 0.962         & 0.051        & 0.968        & 0.954        & 0.062        \\ \hline
FakeAVCeleb               & 0.955        & 0.935         & 0.088        & 0.947        & 0.938        & 0.108        \\ \hline
LAV-DF                    & 0.968        & 0.952         & 0.068        & 0.956        & 0.941        & 0.074        \\ \hline
TVIL                      & 0.932        & 0.912         & 0.118        & 0.913        & 0.890        & 0.145        \\ \hline
\end{tabular}}
\end{table}

The results presented in Table~\ref{table:cross-dataset-results-fakeavceleb} and Table~\ref{table:cross-dataset-results-avdf1m} demonstrate the performance of our MMMS-BA model on intra- and cross datasets when trained on the FakeAVCeleb and AV-Deepfake1M datasets. The model trained and tested on the FakeAVCeleb obtained an AUC and ACC of $0.989$ and $0.979$ respectively. When trained and tested on the AV-Deepfake1M dataset, the model has an AUC and ACC of $0.979$ and $0.968$, respectively. This indicates the model's ability to effectively learn the characteristics of audio-visual manipulations present in the FakeAVCeleb and AV-Deepfake1M datasets.

Table~\ref{table:cross-dataset-results-fakeavceleb} shows that the cross-dataset evaluations have a performance landscape. When trained on FakeAVCeleb and tested on AV-Deepfake1M, LAV-DF, and TVIL. On average, there is a decrease of only $0.053$ and $0.076$ AUC and ACC, respectively, over intra-dataset evaluation.
%there is a decrease of $0.081$ and $0.158$  for the AV-Deepfake1M dataset, $0.032$ and $0.047$  for LAV-DF, $0.048$ and $0.025$ for TVIL datasets.  
For the AV-Deepfake1M dataset, the model obtained the lowest performance with an AUC and ACC of $0.909$ and $0.893$, respectively. This is expected due to the differences in manipulation techniques across datasets. The model maintains relatively high performance when tested on LAV-DF and TVIL. This could be because of the similar manipulation technique in the LAV-DF dataset. While the TVIL dataset has inpainting-based manipulation providing visual cues for detection.

When trained on AV-Deepfake1M and tested on FakeAVCeleb, LAV-DF, and TVIL datasets, a decrease of only $0.027$ and $0.029$ AUC and ACC, respectively, is recorded over the intra-dataset evaluation.
%in performance is observed. In terms of AUC and ACC over the intra dataset evaluation, on average there is a decrease of $0.027$ and $0.029$.
%$0.024$ and $0.021$ for the FakeAVCeleb dataset, $0.011$ and $0.012$ for LAV-DF, $0.047$ and $0.055$ for TVIL datasets .
The model obtained the lowest performance with an AUC of $0.932$ and ACC of $0.913$ on the TVIL dataset. This is expected due to the differences in manipulation techniques across datasets.
The model obtained the best performance on the LAV-DF dataset with AUC and ACC of $0.968$ and $0.956$, respectively. This is because both AV-Deepfake1M and LAV-DF use content-driven manipulation techniques (see Table~\ref{tab:datasets}).

\begin{table}
\centering
\caption{Comparison of MMMS-BA with published audio-visual deepfake detection approaches trained and tested on the FakeAVCeleb dataset.}
\label{table:comparison-results}
\scalebox{0.87}
{
\begin{tabular}{l|c|c}
\hline
\multicolumn{1}{c|}{\textbf{Model}} & \textbf{AUC} & \textbf{ACC} \\ \hline
MIS-AVoiDD~\cite{sree2023mis} & 0.973 & 0.962 \\ \hline
%Modality Agnostic Deepfake Detection~\cite{yu2023modality} & 0.997 & 0.999 \\ \hline
{\begin{tabular}[c]{@{}l@{}}Unsupervised Multi-modal\\ Deepfake Detection~\cite{tian2023unsupervised}\end{tabular}}~ & 0.968 & - \\ \hline
Audio-visual person-of-interest~\cite{cozzolino2023audio} & 0.946 & 0.850 \\ \hline
Audio-visual anomaly detection~\cite{feng2023self} & 0.942 & 0.945 \\ \hline
Multi-modal contrastive learning~\cite{liu2023mcl} & 0.978 & 0.965 \\ \hline
PVAS-MDD~\cite{yu2023pvass} & 0.965 & 0.948 \\ \hline
Novel Smart Deepfake Detection~\cite{elpeltagy2023novel} & 0.954 & 0.960 \\ \hline
Multimodaltrace~\cite{raza2023multimodaltrace} & 0.929 & - \\ \hline
Hearing and seeing abnormality~\cite{sung2023hearing} & 0.944 & - \\ \hline
NPVforensics~\cite{chen2023npvforensics} & 0.925 & - \\ \hline
{\begin{tabular}[c]{@{}l@{}}\textbf{Our-MMMS-BA} (Visual, Audio,\\ and Lip Sequence)\end{tabular}} & \textbf{0.989} & \textbf{0.979} \\ \hline
% MMMS-BA (Visual, audio, and lip sequence) & \textbf{0.989} & \textbf{0.979} \\ \hline
\end{tabular}}
\end{table}

% In cross-dataset evaluations, our models continue to exhibit robust performance, particularly noteworthy when trained on the expansive AV-Deepfake1M dataset and tested across FakeAVCeleb, LAV-DF, and TVIL datasets. This robustness is indicative of the models' generalization capabilities, a critical attribute considering the ever-evolving landscape of deepfake generation techniques.

In \textbf{summary}, the experimental results for deepfake detection suggest a high generalization and efficacy of the MMMS-BA model. On the intra-dataset evaluation of the model on FakeAVCeleb and AV-Deepfake1M datasets, an average AUC and ACC of $0.984$ and $0.973$, respectively, is obtained. The high performance of the model on the FakeAVCeleb dataset also confirms that our model obtains efficient detection performance even when one of the modalities (audio or visual) is manipulated, as available in the FakeAVCeleb dataset.
On cross-dataset evaluation of the model, an average AUC and ACC of $0.944$ and $0.919$, respectively, are obtained, suggesting the high performance and generalizability of our proposed MMMS-BA model. 

%With the intra-dataset evaluation of FakeAVCeleb and AV-Deepfake1M datasets, the model has obtained an AUC of $0.989$ and ACC of $0.979$ for FakeAVCeleb and an AUC of $0.979$ and ACC of $0.968$ for AV-Deepfake1M. For the intra-dataset evaluation on average, there is a performance drop  
%of $0.053$ and $0.076$ in terms of AUC and ACC. Over the intra-dataset evaluation on the AV-Deepfake1M dataset, there is a decrease of $0.027$ and $0.029$ in terms of AUC and ACC. 

\subsection{Comparison with the Published Audio-Visual Deepfake Detectors}

The comparison shown in Table~\ref{table:comparison-results} highlights the competitive landscape of deepfake detection models trained and evaluated on the FakeAVCeleb dataset. As can be seen in Table~\ref{table:comparison-results}, the MMMS-BA model, which uses the full facial image, audio, and lip modality, outperforms all existing audio-visual deepfake detectors by obtaining an AUC of $0.989$ and an ACC of $0.979$. In particular, our MMMS-BA model outperforms NPVForensics~\cite{chen2023npvforensics} which also utilizes face, audio, and lip movement data. NPVForensics is based on mining the correlation between non-critical phonemes and visemes using a Swin Transformer and cross-modal fusion, but it achieves lower performance over MMMS-BA due to its limited ability to capture temporal dependencies and interactions across multiple sequences.
%The NPVForensics model achieves lower performance than the MMMS-BA framework because it does not fully leverage multi-sequence attention, limiting its ability to capture temporal dependencies and interactions across multiple sequences. 
%The reason is a better utilization of the available information from modalities with contextual cross-modal attention.  
%Our model, which leverages visual, audio, and lip modality data, obtained an AUC of $0.989$ and an ACC of $0.979$. This performance positions our model among the top-performing frameworks,  the efficacy of integrating multi-modal data and attention mechanisms.
% The Modality Agnostic Deepfake Detection~\cite{yu2023modality} model sets a benchmark with nearly perfect AUC and ACC scores of $0.997$ and $0.999$, respectively. This exceptional performance suggests a highly effective approach to modality integration, potentially less generalizable across diverse deepfake generation techniques.
The Unsupervised Cross-Modal Inconsistencies~\cite{tian2023unsupervised} model obtained the second-best performance, with an AUC of $0.968$. This also suggests the potential of leveraging motion inconsistencies between modalities for deepfake detection. 
%However, the absence of reported ACC underscores the challenges in obtaining high accuracy. 
% present a spectrum of performances, with AUC ranging from $0.925$ to $0.978$. These variations reflect the diversity in methodologies, from unsupervised learning approaches to those heavily reliant on specific modalities. Notably, the model ~\cite{cozzolino2023audio} shows a significant disparity between AUC ($0.946$) and ACC ($0.850$), highlighting potential issues in the model and balancing of datasets.

In \textbf{summary}, MMMS-BA  outperforms all recently published work based on integrating audio, visual, and lip-movement data with an average performance increment of $0.0341$ and $3.47$\% in AUC and ACC, respectively. Thus, obtaining state-of-the-art performance.
%~\cite{cozzolino2023audio}, ~\cite{feng2023self}, ~\cite{liu2023mcl}, ~\cite{yu2023pvass}, ~\cite{elpeltagy2023novel}, ~\cite{raza2023multimodaltrace}, ~\cite{sung2023hearing}, and ~\cite{chen2023npvforensics} with an average performance increment of $0.0341$ and $3.47$\% in terms of AUC and ACC. Overall, compared to the existing multi-modal methods, our approach obtains state-of-the-art performance.

\begin{table*}[ht]
\centering

\caption{Comparison between MMMS-BA and other methods on LAV-DF, TVIL, and AV-Deepfake1M datasets. Bold faces represent the best performance. (AP- Average Precision and AR- Average Recall).}
\scalebox{0.87}{
\begin{tabular}{@{}llccccccc@{}}
 \hline
\textbf{Testing Dataset} & \textbf{Method} & \textbf{AP@0.5 }& \textbf{AP@0.75 }& \textbf{AP@0.95 }& \textbf{AR@10} & \textbf{AR@20 }& \textbf{AR@50} & \textbf{AR@100} \\ \hline

\multirow{3}{*}{AV-Deepfake1M}
& Enc-Dec  & 06.23 & 0.08 & 0.00 & 11.45 & 15.79 &23.75  & 31.71 \\
& ActionFormer~\cite{zhang2022actionformer}  & 36.08 & 12.01 & 00.16 &  26.60& 27.00 & 27.11 & - \\
& BA-TFD+~\cite{cai2023glitch} & 44.42 & 13.64 & 00.03 & 34.67 & 40.37 & 48.86 & - \\
& UMMAFormer~\cite{zhang2023ummaformer} &51.64 & 28.07 & 01.58 & 42.09 & 43.45 & 48.86 & - \\
& \textbf{Our-MMMS-BA}  & \textbf{62.75} & \textbf{35.87} & \textbf{18.37} & \textbf{54.28}& \textbf{55.94} & \textbf{57.49} & \textbf{59.66}  \\ 
\hline
\multirow{3}{*}{LAV-DF}
& Enc-Dec & 12.96 & 0.97 & 0.00 & 18.79& 21.67 & 30.54 & 38.12 \\
& ActionFormer~\cite{zhang2022actionformer} & 85.23 & 59.05 & 00.93 & 76.93 & 77.19 & 77.23 & 77.23 \\
& BA-TFD+~\cite{cai2023glitch} & 96.30 & 84.96 & 04.44 & 78.75 & 79.40 & 80.48 & 81.62 \\
& \textbf{Our-MMMS-BA}  & \textbf{97.56}  & \textbf{95.25}   & \textbf{39.02 } & \textbf{89.42} & \textbf{95.93} & \textbf{93.45}  & \textbf{94.05}  \\ \hline
\multirow{3}{*}{TVIL}
& Enc-Dec  & 23.08 & 08.45 & 05.32 & 17.16& 23.47 & 43.98 & 45.18 \\
& ActionFormer~\cite{zhang2022actionformer}  & 86.27 & 83.03 & 28.17 &84.82 &85.77  &88.10  & 88.49  \\
& BA-TFD+~\cite{cai2023glitch} & 76.90 & 38.50 & 0.25 & 66.90 & 64.08 & 60.77 & 58.42 \\
& UMMAFormer~\cite{zhang2023ummaformer} & 88.68 & \textbf{84.70} & \textbf{62.43} & 87.09 & \textbf{88.21} & 90.43 & 91.16 \\ 
& \textbf{Our-MMMS-BA}  & \textbf{96.87}  & \textbf{81.33 }   & 28.43   & \textbf{88.61}  & 87.83  & \textbf{90.47}  & \textbf{92.91}  \\ \hline
\end{tabular}}
\label{localization}
\end{table*}

% \begin{table}[htbp]
% \centering
% \caption{Localization Results trained on AV-Deepfake-1M and tested on the LAV-DF, TVIL Datasets}
% \label{tab:results}
% \scalebox{0.67}{
% \begin{tabular}{|l|l|l|l|l|l|l|l}
% \hline
% \textbf{Metric \& Dataset} & \textbf{LAV-DF} & \textbf{TVIL} & \textbf{AV-Deepfake-1M} \\ \hline
% \textbf{AP@0.5} & 97.565 & 92.015 &  \\ \hline
% \textbf{AP@0.75} & 95.25 & 71.3 &  \\ \hline
% \textbf{AP@0.95} & 39.025 & 18.95 &  \\ \hline
% \textbf{AR@10} & 89.425 & 75.71 &  \\ \hline
% \textbf{AR@20} & 95.935 & 76.83 &  \\ \hline
% \textbf{AR@50} & 93.45 & 78.47 &  \\ \hline
% \textbf{AR@100} & 94.05 & 79.91 &  \\ \hline
% \end{tabular}}
% \end{table}

\subsection{Deepfake Localization Performance}

Table~\ref{localization} shows the performance evaluation of the encoder-decoder (Enc-Dec), ActionFormer~\cite{zhang2022actionformer}, BA-TFD+~\cite{cai2023glitch}, UMMAFormer~\cite{zhang2023ummaformer} and MMMS-BA based deepfake localization approaches when trained on AV-Deepfake1M and tested on LAV-DF, TVIL, and AV-Deepfake1M datasets. The Enc-Dec approach represents a sequence-to-sequence encoder-decoder-based architecture employed considering audio and visual modalities. For Enc-Dec, audio and visual encoders are used, and the hidden representations from these encoders are concatenated together, followed by two dense layers, and the output classification layer provides sequences.
%and the corresponding timestamps of the fake sequences determine the localized deepfake segments. 
The start and end timestamps of the fake sequences are the fake segments in a video.  The results for methods ActionFormer, BA-TFD+, and UMMAFormer reported in Table~\ref{localization} are taken from the original papers~\cite{cai2023glitch,zhang2023ummaformer,cai2023av}. Since the evaluation is done on the same datasets used in our study, it is a fair comparison. 

The Enc-Dec model obtained the lowest performance across all datasets and metrics. This is expected as it is based on simple feature concatenation from audio-visual streams and does not include any advanced processing techniques like attention mechanism.
%which is expected as it represents a simpler or more conventional approach compared to the others. 
On AV-Deepfake1M, the MMMS-BA model has obtained the best performance with AP@0.5 and AR@50 of $62.75$ and $57.49$, respectively. UMMAFormer is the second best model obtaining AP@0.5 and AR@0.5 of $51.64$ and $48.86$, respectively. MMMS-BA has obtained better performance on the AV-Deepfake1M dataset when compared to the best model in ~\cite{cai2023av}. However, the lower performance observed on the AV-Deepfake1M dataset, although it was used as a training set, underscores the highly realistic fake content generated in a content-driven manner, altering the real transcripts with replace, delete, and insert operations and the corresponding audio-visual modalities accordingly.
%influenced by rapid advancements in content generation.

When evaluated on the LAV-DF dataset, MMMS-BA obtained the best performance with an AP@0.5 and AR@50 of $97.56$  and $93.45$, respectively. BA-TFD+ has obtained the second-best performance with an AP@0.5 and AR@50 of $96.30$  and $80.48$, respectively. 
%BA-TFD+ has a lower performance differential when compared to MMMS-BA at lower threshold levels but at higher thresholds, it has a higher performance differential indicating less precise localization. 
Although BA-TFD+ showed a smaller performance difference compared to MMMS-BA at lower threshold levels, it exhibited a larger performance difference at higher thresholds, indicating a less precise localization.

On the TVIL dataset, the MMMS-BA model demonstrated the best performance with an AP@0.5 of $96.87$ and an AR@50 of $90.47$. The UMMAFormer model followed closely with an AP@0.5 of $88.68$ and an AR@50 of $90.43$, and it achieved the highest scores for AP@0.75 ($84.70$) and AP@0.95 ($62.43$). The Enc-Dec model had the lowest performance with an AP@0.5 of $23.08$ and an AR@50 of $43.98$ due to its ineffective use of the available modality information. 
%The superior performance of MMMS-BA is particularly evident in its high AP and AR scores across most thresholds, indicating precise and reliable localization.

In \textbf{summary}, the MMMS-BA model consistently demonstrated the best localization capability across all datasets. This is due to the utilization of contextual information across the sequences over existing methods, resulting in better robustness in deepfake localization.  

\section{Ablation Study}
\label{ablation}
An ablation study is conducted by varying the attention mechanism in the contextual cross-modal attention block in Figure~\ref{fig:img2b} resulting in two other variations, i.e., Multi-Modal Uni-Sequences - Self-Attention (MMUS-SA) and Multi-Sequence - Self-Attention (MS-SA). We also ablated between the bi-modal combination of the modalities considered, that is, the full face, voice, and lip region. The details given below.

%Our MMMS-BA model applies bi-modal pairwise attention over multiple modalities and sequences (MMMS-BA). The other two variations of the attention mechanism are Multi-Modal Uni-Sequences - Self-Attention (MMUS-SA) and Multi-Sequence - Self-Attention (MS-SA). The following sections provide details on these variations, followed by the performance of the model. We also ablated between the bi-modal combination of the modalities considered, that is, the full face, voice and lip region, details given in the subsection~\ref{varying-modalities}. Note these ablation studies are conducted only for deepfake detection performance.

\subsection{Varying the Attention at Sequence and Modality Level}
\label{varyingattention}
\textbf{Multi-Modal Uni-Sequences - Self Attention (MMUS-SA) Framework}: MMUS-SA framework does not account for information from other sequences at the attention level but utilizes multi-modal information of a single sequence for prediction. For more details on the calculation of attention in the MMUS-SA framework refer to the Appendix, section $2$.

\textbf{Multi-Sequence - Self Attention (MS-SA) Framework}: In the MS-SA framework, we apply self-attention to the sequences of each modality separately and use these for classification. For more details on the calculation of attention in the MMUS-SA framework refer to the Appendix, section $2$.

% \begin{itemize}

% \item \textbf{Multi-Modal Uni-Sequences - Self Attention (MMUS-SA) Framework}: MMUS-SA framework does not account for information from other sequences at the attention level but utilizes multi-modal information of a single sequence for prediction. For more details on the calculation of attention in the MMUS-SA framework refer to the Appendix in Section ~\ref{appendix}.

% \item \textbf{Multi-Sequence - Self Attention (MS-SA) Framework}: In the MS-SA framework, we apply self-attention to the sequences of each modality separately and use these for classification. For more details on the calculation of attention in the MMUS-SA framework refer to the Appendix in Section ~\ref{appendix}. 
% \end{itemize}

%\subsection{Performance of Deepfake Detection with Attention Variations}
Table~\ref{table:fakeavceleb-attention-variations} shows the performance of the MMUS-SA and MS-SA models over MMMS-BA when trained and tested on FakeAVCeleb.  From the results obtained, it is evident that MMMS-BA has obtained the best performance with the lowest EER of $0.029$. The MMUS-SA model has a closer performance with the MMMS-BA variation but has a higher EER. This suggests that incorporating self-attention into a single sequence in multiple modalities achieved a closer AUC, but was unable to obtain a low EER. However, MS-SA has obtained the lowest performance when compared to the other two variations. The reason is that MS-SA does not incorporate the interaction between the modalities for multi-modal manipulation detection.

In \textbf{summary}, MS-SA and MMUS-SA attention methods typically identify important features within a modality over multiple sequences and a single sequence across different modalities, respectively. However, the proposed MMMS-BA model is specifically designed to understand the intricate relationship between different modalities over multiple sequences, leading to a more robust representation of the data and enhanced performance. Consequently, obtaining better performance over MMUS-SA and MS-SA. This confirms the importance of using interactions between both multi-sequences and multi-modalities for deepfake detection.

%The proposed contextual cross-modal attention approach represents a significant advancement over basic attention mechanisms in several key aspects. It enhances the focus on cross-modal relationships by considering the limitations of single-modality attention mechanisms.
% The approach facilitates improved representation learning by considering the context and relationships between different modalities. By integrating information from multiple modalities, the model can learn more comprehensive and representative features, thereby enhancing its ability to discriminate between authentic and manipulated media content.

\begin{table}
\centering
\caption{Performance comparison of MMMS-BA with MMUS-BA and MS-SA models with varied attention mechanism, trained and tested on FakeAVCeleb dataset.}
\label{table:fakeavceleb-attention-variations}
\scalebox{0.83}{
\begin{tabular}{l|c|c|c|c|c|c}
\hline
\textbf{Model} & \textbf{AUC} & \textbf{pAUC} & \textbf{EER} & \textbf{ACC} & \textbf{TPR} & \textbf{FPR} \\ \hline
MMUS-SA           & 0.989       & 0.978         & 0.033        & 0.979        & 0.963        & 0.039        \\ \hline
MS-SA                   & 0.977        & 0.956         & 0.045       & 0.965       & 0.943        & 0.051       \\ \hline
MMMS-BA  & \textbf{0.989}        & \textbf{0.977}         & \textbf{0.029}        & \textbf{0.979 }       & \textbf{0.965}        & \textbf{0.039}        \\ \hline
\end{tabular}}
\end{table}

\subsection{Performance of MMMS-BA on Different Combination of Modalities}
\label{varying-modalities}
In this study, we ablated between the combination of modalities (V-full visual face, L-lip sequence, and A-audio) and evaluated the performance of MMMS-BA when trained and tested on FakeAVCeleb. 
Table~\ref{table:fakeavceleb-modlaity-variations} illustrates the role of different modalities in enhancing detection accuracy. 
As can be seen, the combination of visual and lip (V + L) obtained the lowest performance with an AUC of $0.814$.  This result can be attributed to the absence of audio data, which limits the model to visual information alone. 
The combination of lip and audio (L + A) obtained a performance improvement over (V + L) with an increase in AUC of $0.11$. This enhancement suggests that incorporating audio data alongside lip sequences provides complementary information, enhancing the model's performance.
The combination of visual and audio (V+A) obtained further improvement over (V+L and L+A) with an AUC of $0.955$. This integration demonstrates that including audio features with full-face visual data offers additional insights, thereby boosting detection accuracy beyond the (V + L) and (L + A) combinations. The best performance is obtained by combining (V+L+A) with an average increment in AUC of $0.091$. 

In \textbf{summary}, this analysis suggests that incorporating lip sequence data along with (V+A) contributes significantly to the effectiveness of the model. Full-face images encompass a complex mix of facial features, making it challenging for the model to isolate and differentiate inconsistencies in the lip region effectively. 
%Additionally, full-face images lack explicit information about the location of the lip region, further complicating the detection process. In contrast, the availability of lip sequence modality offers a clear and isolated view of the lips, facilitating the extraction of lip-related features with greater precision. 
Thus, the additional lip modality enriched the information available to the model, ultimately improving its discriminatory power and performance, also supported by the NPVForensics model~\cite{chen2023npvforensics} for audio-visual deepfake detection.

\begin{table}
\centering
\caption{Performance comparison of MMMS-BA with different bi-modal combinations (V-full visual face, L-lip sequence, and A-audio) when trained and tested on the FakeAVCeleb dataset.}
\label{table:fakeavceleb-modlaity-variations}
\scalebox{0.71}{
\begin{tabular}{l|c|c|c|c|c|c}
\hline
\textbf{Model and Modalities} & \textbf{AUC} & \textbf{pAUC} & \textbf{EER} & \textbf{ACC} & \textbf{TPR} & \textbf{FPR} \\ \hline
MMMS-BA  (V+A) & 0.955        & 0.941         & 0.074        & 0.939        & 0.922        & 0.095        \\ \hline
MMMS-BA  (V+L) &   0.814      &     0.798     &    0.263     &    0.736     &    0.818     &      0.345   \\ \hline
MMMS-BA  (L+A) &    0.924    &    0.907     &  0.175      &    0.823     &    0.931     &       0.282  \\ \hline
MMMS-BA (V+L+A)           & \textbf{0.989 }      & \textbf{0.978}         & \textbf{0.033 }       & \textbf{0.979 }       & \textbf{0.963}        & \textbf{0.039 }       \\ \hline
\end{tabular}}
\end{table}

\section{Conclusion and Future Work}
\label{conclusion}

With the rapid evolution of deepfake techniques combining synthesized audio with forged videos, there is an urgent need for robust audio-visual deepfake detection and localization techniques. Current audio-visual deepfake detection approaches relying on fusing audio and visual streams employing basic attention mechanisms, overlook intricate inter-modal relationships crucial for accurate detection. To address this challenge, we introduced the MMMS-BA framework. This framework effectively captures intra- and inter-modal correlations by applying attention to multi-modal multi-sequence representations and learns the contributing features among
them for effective deepfake detection and localization
%To address this, we introduced the MMMS-BA framework, which effectively captures intra- and inter-modal correlations across multi-modalities and multi-sequences in a video. 
Our experimental findings demonstrate that MMMS-BA outperforms existing audio-visual deepfake detectors, achieving SOTA performance in detecting and localizing deepfake segments within videos.

%, to further enhance detection accuracy. 
Given the proliferation of AI-generated content using sophisticated generative models, tackling emerging forms of content manipulation remains a critical challenge. Therefore, as part of future research directions, we will extend our framework to incorporate text analysis along with audio and visual modalities to ensure robust protection against misleading multimedia content. In addition, our model will be adapted to account for missing modalities during the training and inference stage.

\balance

{\small
\bibliographystyle{ieee}
\bibliography{Bookchapter}

\begin{thebibliography}{10}\itemsep=-1pt

\bibitem{8630761}
D.~Afchar, V.~Nozick, J.~Yamagishi, and I.~Echizen.
\newblock Mesonet: a compact facial video forgery detection network.
\newblock In {\em 2018 IEEE International Workshop on Information Forensics and Security (WIFS)}, pages 1--7, 2018.

\bibitem{9360904}
S.~Agarwal, H.~Farid, T.~El-Gaaly, and S.-N. Lim.
\newblock Detecting deep-fake videos from appearance and behavior.
\newblock In {\em 2020 IEEE International Workshop on Information Forensics and Security (WIFS)}, pages 1--6, 2020.

\bibitem{agarwal2020detecting}
S.~Agarwal, H.~Farid, O.~Fried, and M.~Agrawala.
\newblock Detecting deep-fake videos from phoneme-viseme mismatches.
\newblock In {\em Proceedings of the IEEE/CVF conference on computer vision and pattern recognition workshops}, pages 660--661, 2020.

\bibitem{agarwal2023watch}
S.~Agarwal, L.~Hu, E.~Ng, T.~Darrell, H.~Li, and A.~Rohrbach.
\newblock Watch those words: Video falsification detection using word-conditioned facial motion.
\newblock In {\em Proceedings of the IEEE/CVF Winter Conference on Applications of Computer Vision}, pages 4710--4719, 2023.

\bibitem{asha2024defensive}
S.~Asha, P.~Vinod, and V.~G. Menon.
\newblock A defensive attention mechanism to detect deepfake content across multiple modalities.
\newblock {\em Multimedia Systems}, 30(1):56, 2024.

\bibitem{bagchi2021hear}
A.~Bagchi, J.~Mahmood, D.~Fernandes, and R.~K. Sarvadevabhatla.
\newblock Hear me out: Fusional approaches for audio augmented temporal action localization.
\newblock {\em arXiv preprint arXiv:2106.14118}, 2021.

\bibitem{bodla2017soft}
N.~Bodla, B.~Singh, R.~Chellappa, and L.~S. Davis.
\newblock Soft-nms--improving object detection with one line of code.
\newblock In {\em Proceedings of the IEEE international conference on computer vision}, pages 5561--5569, 2017.

\bibitem{cai2023av}
Z.~Cai, S.~Ghosh, A.~P. Adatia, M.~Hayat, A.~Dhall, and K.~Stefanov.
\newblock Av-deepfake1m: A large-scale llm-driven audio-visual deepfake dataset.
\newblock {\em arXiv preprint arXiv:2311.15308}, 2023.

\bibitem{cai2023glitch}
Z.~Cai, S.~Ghosh, T.~Gedeon, A.~Dhall, K.~Stefanov, and M.~Hayat.
\newblock " glitch in the matrix!": A large scale benchmark for content driven audio-visual forgery detection and localization.
\newblock {\em arXiv preprint arXiv:2305.01979}, 2023.

\bibitem{cai2022you}
Z.~Cai, K.~Stefanov, A.~Dhall, and M.~Hayat.
\newblock Do you really mean that? content driven audio-visual deepfake dataset and multimodal method for temporal forgery localization.
\newblock In {\em 2022 International Conference on Digital Image Computing: Techniques and Applications (DICTA)}, pages 1--10. IEEE, 2022.

\bibitem{chen2023npvforensics}
Y.~Chen, Y.~Yu, R.~Ni, Y.~Zhao, and H.~Li.
\newblock Npvforensics: Jointing non-critical phonemes and visemes for deepfake detection.
\newblock {\em arXiv preprint arXiv:2306.06885}, 2023.

\bibitem{cheng2023voice}
H.~Cheng, Y.~Guo, T.~Wang, Q.~Li, X.~Chang, and L.~Nie.
\newblock Voice-face homogeneity tells deepfake.
\newblock {\em ACM Transactions on Multimedia Computing, Communications and Applications}, 20(3):1--22, 2023.

\bibitem{chintha2020recurrent}
A.~Chintha, B.~Thai, S.~J. Sohrawardi, K.~Bhatt, A.~Hickerson, M.~Wright, and R.~Ptucha.
\newblock Recurrent convolutional structures for audio spoof and video deepfake detection.
\newblock {\em IEEE Journal of Selected Topics in Signal Processing}, 14(5):1024--1037, 2020.

\bibitem{Choi_2018_CVPR}
Y.~Choi, M.~Choi, M.~Kim, J.-W. Ha, S.~Kim, and J.~Choo.
\newblock Stargan: Unified generative adversarial networks for multi-domain image-to-image translation.
\newblock In {\em Proceedings of the IEEE Conference on Computer Vision and Pattern Recognition (CVPR)}, June 2018.

\bibitem{10.1145/3394171.3413700}
K.~Chugh, P.~Gupta, A.~Dhall, and R.~Subramanian.
\newblock Not made for each other- audio-visual dissonance-based deepfake detection and localization.
\newblock In {\em Proceedings of the 28th ACM International Conference on Multimedia}, MM '20, page 439–447, New York, NY, USA, 2020. Association for Computing Machinery.

\bibitem{chung18b_interspeech}
J.~S. Chung, A.~Nagrani, and A.~Zisserman.
\newblock {VoxCeleb2: Deep Speaker Recognition}.
\newblock In {\em Proc. Interspeech 2018}, pages 1086--1090, 2018.

\bibitem{citron}
D.~Citron.
\newblock How deepfakes undermine truth and threaten democracy.

\bibitem{cozzolino2023audio}
D.~Cozzolino, A.~Pianese, M.~Nie{\ss}ner, and L.~Verdoliva.
\newblock Audio-visual person-of-interest deepfake detection.
\newblock In {\em Proceedings of the IEEE/CVF conference on computer vision and pattern recognition}, pages 943--952, 2023.

\bibitem{elpeltagy2023novel}
M.~Elpeltagy, A.~Ismail, M.~S. Zaki, and K.~Eldahshan.
\newblock A novel smart deepfake video detection system.
\newblock {\em International Journal of Advanced Computer Science and Applications}, 14(1), 2023.

\bibitem{feng2023self}
C.~Feng, Z.~Chen, and A.~Owens.
\newblock Self-supervised video forensics by audio-visual anomaly detection.
\newblock In {\em Proceedings of the IEEE/CVF Conference on Computer Vision and Pattern Recognition}, pages 10491--10503, 2023.

\bibitem{gao2018ctap}
J.~Gao, K.~Chen, and R.~Nevatia.
\newblock Ctap: Complementary temporal action proposal generation.
\newblock In {\em Proceedings of the European conference on computer vision (ECCV)}, pages 68--83, 2018.

\bibitem{gao2017turn}
J.~Gao, Z.~Yang, K.~Chen, C.~Sun, and R.~Nevatia.
\newblock Turn tap: Temporal unit regression network for temporal action proposals.
\newblock In {\em Proceedings of the IEEE international conference on computer vision}, pages 3628--3636, 2017.

\bibitem{9578910}
A.~Haliassos, K.~Vougioukas, S.~Petridis, and M.~Pantic.
\newblock Lips don't lie: A generalisable and robust approach to face forgery detection.
\newblock In {\em 2021 IEEE/CVF Conference on Computer Vision and Pattern Recognition (CVPR)}, pages 5037--5047, 2021.

\bibitem{hamza2022deepfake}
A.~Hamza, A.~R.~R. Javed, F.~Iqbal, N.~Kryvinska, A.~S. Almadhor, Z.~Jalil, and R.~Borghol.
\newblock Deepfake audio detection via mfcc features using machine learning.
\newblock {\em IEEE Access}, 10:134018--134028, 2022.

\bibitem{Hwang2020}
T.~Hwang.
\newblock Deepfakes: A grounded threat assessment.
\newblock Technical report, Georgetown University, July 2020.

\bibitem{ILYAS2023110124}
H.~Ilyas, A.~Javed, and K.~M. Malik.
\newblock Avfakenet: A unified end-to-end dense swin transformer deep learning model for audio–visual​ deepfakes detection.
\newblock {\em Applied Soft Computing}, 136:110124, 2023.

\bibitem{katamneni_nadimpalli_rattani_2023}
V.~S. Katamneni, A.~V. Nadimpalli, and A.~Rattani.
\newblock Demographic fairness and accountability of audio and video-based unimodal and bi-modal deepfake detectors.
\newblock In T.~Bourlai, editor, {\em Face Recognition Across the Imaging Spectrum (FRAIS)}. Springer, 2023.

\bibitem{10.1145/3476099.3484315}
H.~Khalid, M.~Kim, S.~Tariq, and S.~S. Woo.
\newblock Evaluation of an audio-video multimodal deepfake dataset using unimodal and multimodal detectors.
\newblock In {\em Proceedings of the 1st Workshop on Synthetic Multimedia - Audiovisual Deepfake Generation and Detection}, ADGD '21, page 7–15, New York, NY, USA, 2021. Association for Computing Machinery.

\bibitem{khalid2021fakeavceleb}
H.~Khalid, S.~Tariq, M.~Kim, and S.~S. Woo.
\newblock Fakeavceleb: A novel audio-video multimodal deepfake dataset.
\newblock {\em arXiv preprint arXiv:2108.05080}, 2021.

\bibitem{DBLP:journals/corr/abs-2108-05080}
H.~Khalid, S.~Tariq, and S.~S. Woo.
\newblock Fakeavceleb: {A} novel audio-video multimodal deepfake dataset.
\newblock {\em CoRR}, abs/2108.05080, 2021.

\bibitem{lewis2020deepfake}
J.~K. Lewis, I.~E. Toubal, H.~Chen, V.~Sandesera, M.~Lomnitz, Z.~Hampel-Arias, C.~Prasad, and K.~Palaniappan.
\newblock Deepfake video detection based on spatial, spectral, and temporal inconsistencies using multimodal deep learning.
\newblock In {\em 2020 IEEE Applied Imagery Pattern Recognition Workshop (AIPR)}, pages 1--9. IEEE, 2020.

\bibitem{9157215}
L.~Li, J.~Bao, T.~Zhang, H.~Yang, D.~Chen, F.~Wen, and B.~Guo.
\newblock Face x-ray for more general face forgery detection.
\newblock In {\em 2020 IEEE/CVF Conference on Computer Vision and Pattern Recognition (CVPR)}, pages 5000--5009, 2020.

\bibitem{Li_2019_CVPR_Workshops}
Y.~Li and S.~Lyu.
\newblock Exposing deepfake videos by detecting face warping artifacts.
\newblock In {\em Proceedings of the IEEE/CVF Conference on Computer Vision and Pattern Recognition (CVPR) Workshops}, June 2019.

\bibitem{lin2017focal}
T.-Y. Lin, P.~Goyal, R.~Girshick, K.~He, and P.~Doll{\'a}r.
\newblock Focal loss for dense object detection.
\newblock In {\em Proceedings of the IEEE international conference on computer vision}, pages 2980--2988, 2017.

\bibitem{liu2023magnifying}
X.~Liu, Y.~Yu, X.~Li, and Y.~Zhao.
\newblock Magnifying multimodal forgery clues for deepfake detection.
\newblock {\em Signal Processing: Image Communication}, 118:117010, 2023.

\bibitem{liu2023mcl}
X.~Liu, Y.~Yu, X.~Li, and Y.~Zhao.
\newblock Mcl: multimodal contrastive learning for deepfake detection.
\newblock {\em IEEE Transactions on Circuits and Systems for Video Technology}, 2023.

\bibitem{masood2023attention}
M.~Masood, A.~Javed, and A.~Irtaza.
\newblock Attention-based multimodal learning framework for generalized audio-visual deepfake detection.
\newblock {\em PREPRINT Research Square : rs.3.rs-3415144/v1}, 2023.

\bibitem{10.1145/3394171.3413570}
T.~Mittal, U.~Bhattacharya, R.~Chandra, A.~Bera, and D.~Manocha.
\newblock Emotions don't lie: An audio-visual deepfake detection method using affective cues.
\newblock In {\em Proceedings of the 28th ACM International Conference on Multimedia}, MM '20, page 2823–2832, New York, NY, USA, 2020. Association for Computing Machinery.

\bibitem{journals/corr/abs-1003-4083}
L.~Muda, M.~Begam, and I.~Elamvazuthi.
\newblock Voice recognition algorithms using mel frequency cepstral coefficient (mfcc) and dynamic time warping (dtw) techniques.
\newblock {\em CoRR}, abs/1003.4083, 2010.

\bibitem{Nadimpalli2022GBDFGB}
A.~V. Nadimpalli and A.~Rattani.
\newblock Gbdf: Gender balanced deepfake dataset towards fair deepfake detection.
\newblock {\em ArXiv}, abs/2207.10246, 2022.

\bibitem{10.1145/3625547}
A.~V. Nadimpalli and A.~Rattani.
\newblock Proactive deepfake detection using gan-based visible watermarking.
\newblock {\em ACM Trans. Multimedia Comput. Commun. Appl.}, Sep 2023.

\bibitem{nekadi2020siamese}
R.~Nekadi.
\newblock {\em Siamese Network-based Multi-modal Deepfake Detection}.
\newblock University of Missouri-Kansas City, 2020.

\bibitem{Nguyen2019DeepLF}
T.~T. Nguyen, C.~M. Nguyen, D.~Nguyen, D.~T. Nguyen, and S.~Nahavandi.
\newblock Deep learning for deepfakes creation and detection.
\newblock {\em ArXiv}, abs/1909.11573, 2019.

\bibitem{Nirkin_2019_ICCV}
Y.~Nirkin, Y.~Keller, and T.~Hassner.
\newblock Fsgan: Subject agnostic face swapping and reenactment.
\newblock In {\em Proceedings of the IEEE/CVF International Conference on Computer Vision (ICCV)}, October 2019.

\bibitem{oorloff2024avff}
T.~Oorloff, S.~Koppisetti, N.~Bonettini, D.~Solanki, B.~Colman, Y.~Yacoob, A.~Shahriyari, and G.~Bharaj.
\newblock Avff: Audio-visual feature fusion for video deepfake detection.
\newblock {\em arXiv e-prints}, pages arXiv--2406, 2024.

\bibitem{pianese2022deepfake}
A.~Pianese, D.~Cozzolino, G.~Poggi, and L.~Verdoliva.
\newblock Deepfake audio detection by speaker verification.
\newblock In {\em 2022 IEEE International Workshop on Information Forensics and Security (WIFS)}, pages 1--6. IEEE, 2022.

\bibitem{9350195}
Y.~Qian, Z.~Chen, and S.~Wang.
\newblock Audio-visual deep neural network for robust person verification.
\newblock {\em IEEE/ACM Transactions on Audio, Speech, and Language Processing}, 29:1079--1092, 2021.

\bibitem{raza2023multimodaltrace}
M.~A. Raza and K.~M. Malik.
\newblock Multimodaltrace: Deepfake detection using audiovisual representation learning.
\newblock In {\em Proceedings of the IEEE/CVF Conference on Computer Vision and Pattern Recognition}, pages 993--1000, 2023.

\bibitem{rezatofighi2019generalized}
H.~Rezatofighi, N.~Tsoi, J.~Gwak, A.~Sadeghian, I.~Reid, and S.~Savarese.
\newblock Generalized intersection over union: A metric and a loss for bounding box regression.
\newblock In {\em Proceedings of the IEEE/CVF conference on computer vision and pattern recognition}, pages 658--666, 2019.

\bibitem{9010912}
A.~Rössler, D.~Cozzolino, L.~Verdoliva, C.~Riess, J.~Thies, and M.~Niessner.
\newblock Faceforensics++: Learning to detect manipulated facial images.
\newblock In {\em 2019 IEEE/CVF International Conference on Computer Vision (ICCV)}, pages 1--11, 2019.

\bibitem{salvi2023timit}
D.~Salvi, B.~Hosler, P.~Bestagini, M.~C. Stamm, and S.~Tubaro.
\newblock Timit-tts: A text-to-speech dataset for multimodal synthetic media detection.
\newblock {\em IEEE Access}, 2023.

\bibitem{shao2024detecting}
R.~Shao, T.~Wu, J.~Wu, L.~Nie, and Z.~Liu.
\newblock Detecting and grounding multi-modal media manipulation and beyond.
\newblock {\em IEEE Transactions on Pattern Analysis and Machine Intelligence}, 2024.

\bibitem{sree2023mis}
V.~Sree~Katamneni and A.~Rattani.
\newblock Mis-avoidd: Modality invariant and specific representation for audio-visual deepfake detection.
\newblock {\em arXiv e-prints}, pages arXiv--2310, 2023.

\bibitem{sun2023ai}
C.~Sun, S.~Jia, S.~Hou, and S.~Lyu.
\newblock Ai-synthesized voice detection using neural vocoder artifacts.
\newblock {\em arXiv preprint arXiv:2304.13085}, 2023.

\bibitem{10095247}
C.-S. Sung, J.-C. Chen, and C.-S. Chen.
\newblock Hearing and seeing abnormality: Self-supervised audio-visual mutual learning for deepfake detection.
\newblock In {\em ICASSP 2023 - 2023 IEEE International Conference on Acoustics, Speech and Signal Processing (ICASSP)}, pages 1--5, 2023.

\bibitem{sung2023hearing}
C.-S. Sung, J.-C. Chen, and C.-S. Chen.
\newblock Hearing and seeing abnormality: Self-supervised audio-visual mutual learning for deepfake detection.
\newblock In {\em ICASSP 2023-2023 IEEE International Conference on Acoustics, Speech and Signal Processing (ICASSP)}, pages 1--5. IEEE, 2023.

\bibitem{tian2023unsupervised}
M.~Tian, M.~Khayatkhoei, J.~Mathai, and W.~AbdAlmageed.
\newblock Unsupervised multimodal deepfake detection using intra-and cross-modal inconsistencies.
\newblock {\em arXiv preprint arXiv:2311.17088}, 2023.

\bibitem{Tolosana2020deepfakesAB}
R.~Tolosana, R.~Vera-Rodr{\'i}guez, J.~Fierrez, A.~Morales, and J.~Ortega-Garcia.
\newblock Deepfakes and beyond: A survey of face manipulation and fake detection.
\newblock {\em Inf. Fusion}, 64:131--148, 2020.

\bibitem{wang2022audio}
G.~Wang, P.~Zhang, L.~Xie, W.~Huang, Y.~Zha, and Y.~Zhang.
\newblock An audio-visual attention based multimodal network for fake talking face videos detection.
\newblock {\em arXiv preprint arXiv:2203.05178}, 2022.

\bibitem{xu2018youtube}
N.~Xu, L.~Yang, Y.~Fan, J.~Yang, D.~Yue, Y.~Liang, B.~Price, S.~Cohen, and T.~Huang.
\newblock Youtube-vos: Sequence-to-sequence video object segmentation.
\newblock In {\em Proceedings of the European conference on computer vision (ECCV)}, pages 585--601, 2018.

\bibitem{Xu_2018_CVPR}
T.~Xu, P.~Zhang, Q.~Huang, H.~Zhang, Z.~Gan, X.~Huang, and X.~He.
\newblock Attngan: Fine-grained text to image generation with attentional generative adversarial networks.
\newblock In {\em Proceedings of the IEEE Conference on Computer Vision and Pattern Recognition (CVPR)}, June 2018.

\bibitem{yu2023pvass}
Y.~Yu, X.~Liu, R.~Ni, S.~Yang, Y.~Zhao, and A.~C. Kot.
\newblock Pvass-mdd: predictive visual-audio alignment self-supervision for multimodal deepfake detection.
\newblock {\em IEEE Transactions on Circuits and Systems for Video Technology}, 2023.

\bibitem{zhang2022actionformer}
C.-L. Zhang, J.~Wu, and Y.~Li.
\newblock Actionformer: Localizing moments of actions with transformers.
\newblock In {\em European Conference on Computer Vision}, pages 492--510. Springer, 2022.

\bibitem{zhang2023ummaformer}
R.~Zhang, H.~Wang, M.~Du, H.~Liu, Y.~Zhou, and Q.~Zeng.
\newblock Ummaformer: A universal multimodal-adaptive transformer framework for temporal forgery localization.
\newblock In {\em Proceedings of the 31st ACM International Conference on Multimedia}, pages 8749--8759, 2023.

\bibitem{zhang2022deepfake}
T.~Zhang.
\newblock Deepfake generation and detection, a survey.
\newblock {\em Multimedia Tools and Applications}, 81(5):6259--6276, 2022.

\bibitem{zhang2021deepfake}
Y.~Zhang, J.~Zhan, W.~Jiang, and Z.~Fan.
\newblock Deepfake detection based on incompatibility between multiple modes.
\newblock In {\em 2021 International Conference on Intelligent Technology and Embedded Systems (ICITES)}, pages 1--7. IEEE, 2021.

\bibitem{zhao2023fine}
L.~Zhao, M.~Zhang, H.~Ding, and X.~Cui.
\newblock Fine-grained deepfake detection based on cross-modality attention.
\newblock {\em Neural Computing and Applications}, pages 1--14, 2023.

\bibitem{zhou2021joint}
Y.~Zhou and S.-N. Lim.
\newblock Joint audio-visual deepfake detection.
\newblock In {\em Proceedings of the IEEE/CVF International Conference on Computer Vision}, pages 14800--14809, 2021.

\end{thebibliography}
}
\section{Supplementary Material}
\begin{figure*}
\centering
    \includegraphics[width=0.65\linewidth]{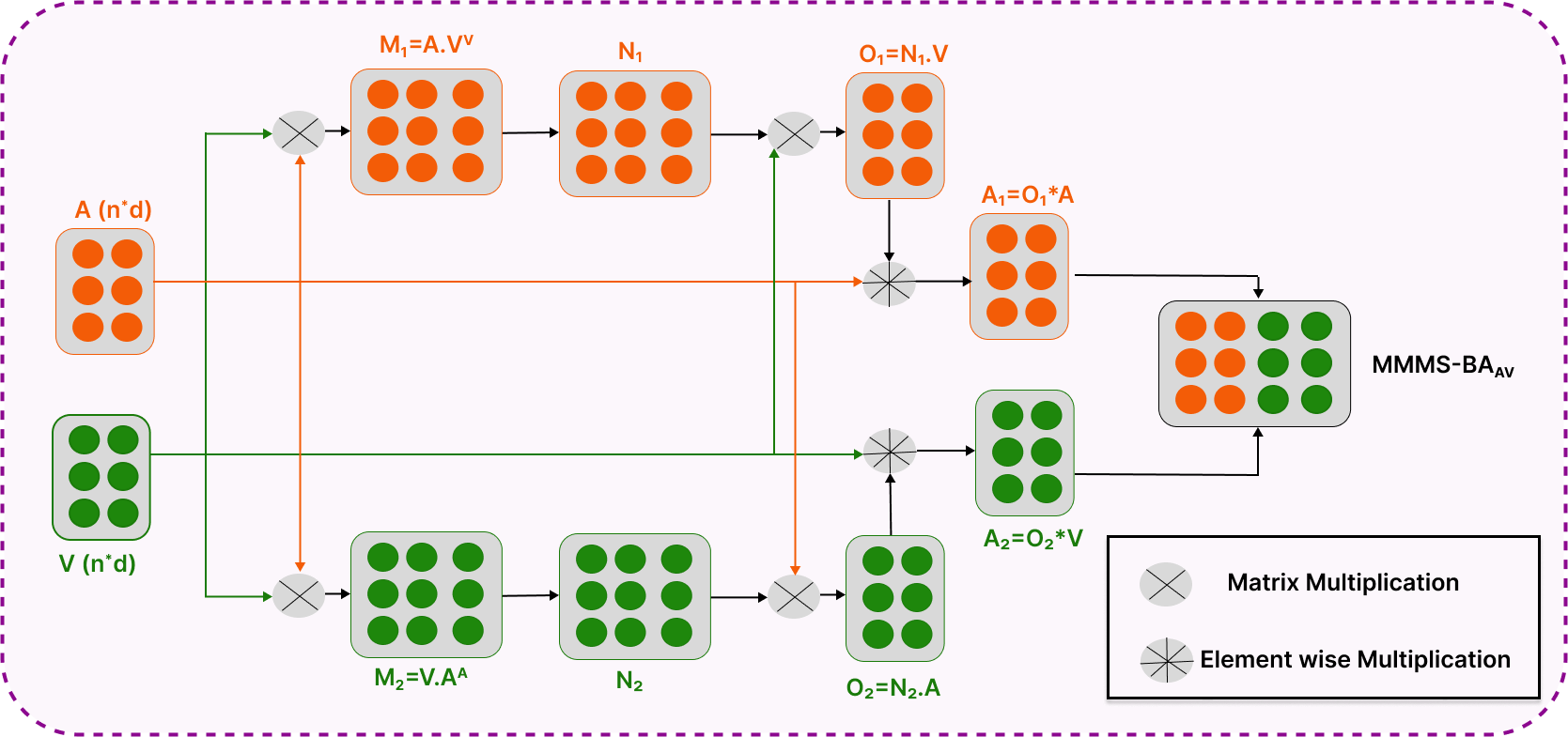}
    \caption{Multi-Modal Multi-Sequence Attention computation of Audio and Full Visual Face Modalities ($MMMS\text{-}BA_{AV}$)}
    \label{fig:img3}
\end{figure*}
\subsection{Dataset Details}
Evaluation of the proposed method is conducted on publicly available audio-visual AV-Deepfake1M~\cite{cai2023av}, FakeAVCeleb~\cite{DBLP:journals/corr/abs-2108-05080}, LAV-DF~\cite{cai2022you}, and TVIL~\cite{zhang2023ummaformer} deepfake datasets mentioned in Section $4.1$ of the main paper. Further, details on these datasets are given below.

\begin{itemize}
    \item \noindent \textbf{FakeAVCeleb:} The FakeAVCeleb dataset~\cite{DBLP:journals/corr/abs-2108-05080} is a collection of videos with audio and visual manipulations of celebrities that have been generated using various deepfake techniques. The dataset is created by selecting videos from the VoxCeleb2~\cite{chung18b_interspeech} dataset, featuring $500$ celebrities. 
The videos in this dataset are clean, featuring only one person's frontal face without any occlusion. 
The dataset is well-balanced and annotated in terms of gender, race, geography, and visual and audio manipulations, making it useful for training deep learning models that can generalize well on unseen test sets.  
We chose this dataset for our multimodal detection experiments because it contains both audio and visual manipulations, as well as a variety of deep-fake generation techniques. 

\item \noindent \textbf{{\textbf{LAV-DF Dataset}}}
\label{lavdf}
The Localized Audio Visual DeepFake (LAV-DF)~\cite{cai2022you} dataset emerges as a critical asset in deepfake detection, particularly for benchmarking methods to detect and localize manipulated segments within videos. 
Comprising 136,304 videos across 153 unique identities, the dataset offers a diverse collection that includes 36,431 real videos alongside 99,873 videos embedded with fake segments. For a complete evaluation, LAV-DF is divided into three identity-independent subsets: training (78,703 videos), validation (31,501 videos), and testing (26,100 videos), each offering a distinct set of identities to ensure an unbiased assessment.

\item \noindent \textbf{{\textbf{TVIL Dataset}}}
\label{tvil}
In response to the increasing challenges posed by advanced AI-generated content (AIGC) technologies, the TVIL dataset~\cite{zhang2023ummaformer} has been synthesized as a new benchmark aimed at locating video inpainting segments. 
% While existing benchmarks predominantly focus on facial or speech content forgeries, TVIL broadens the spectrum by targeting various types of inpainting forgery prevalent in sequential images or videos.
% This initiative seeks to mitigate the potential harm inflicted by the spreading of misleading information through video and audio content. While existing benchmarks predominantly focus on facial or speech content forgeries, TVIL broadens the spectrum by targeting various types of inpainting forgery prevalent in sequential images or videos.
Constructed upon the foundation of YouTubeVOS 2018 ~\cite{xu2018youtube}, which aggregates over 4,000 videos from YouTube, the TVIL dataset leverages one of the most prolific platforms for video content as its base. This choice is strategic, considering YouTube's prominence in content generation and the spread of misinformation. By synthesizing a dataset rooted in YouTube videos, TVIL is poised to offer a robust evaluation framework that defends against misinformation and catalyzes new research directions in the fight against digital content forgery.

\item \noindent \textbf{{\textbf{AV-Deepfake1M Dataset}}}
\label{avdeepfake1m}
The AV-Deepfake1M~\cite{cai2023av} dataset stands as a pioneering contribution to the field of audio-visual deepfake detection, encompassing an extensive compilation of 1,886 hours of audio-visual data curated using content-driven manipulation techniques.
from 2,068 unique subjects set against diverse background environments. 
This dataset is distinguished in facilitating the development of detection algorithms capable of navigating the landscape of content-driven audio-visual manipulations.
\end{itemize}

\subsection{On Varying Attention Mechanism}
This section provides more details on the calculation of attention in MMUS-SA and MS-SA variations discussed in Section 5.1 of the main paper.
\begin{itemize}

\item \textbf{Multi-Modal Uni-Sequences - Self Attention (MMUS-SA) Framework}: MMUS-SA framework does not account for information from other sequences at the attention level, rather it utilizes multi-modal information of a single sequence for prediction. 
% For a video $D$ having ‘$N$’ sequences, ‘$N$’ separate attention blocks are needed, where each block computes the self-attention over multi-modal information of a single sequence. Let $X_{p}$ be the information matrix of the $p^{th}$ sequence where the three ‘$r$’ dimensional rows are the outputs of the dense layers for the three modalities. The attention matrix $A_{p}$ is computed separately for $p = 1^{st}$, $2^{nd}$, ... $N^{th}$ sequences. Finally, for each sequence $p$, $A_{p}$ and $X_{p}$ are concatenated and passed to the output layer for classification.
For a video $D$ having ‘$N$’ sequences, ‘$N$’ separate attention blocks are needed, where each block computes the self-attention over multi-modal information of a single sequence. Let $X_{p}$ be the information matrix of the $p^{th}$ sequence where the three ‘$r$’ dimensional rows are the outputs of the dense layers for the three modalities. The attention matrix $A_{p}$ is computed separately for $p = 1^{st}$, $2^{nd}$, ... $N^{th}$ sequences. Finally, for each sequence $p$, $A_{p}$ and $X_{p}$ are concatenated and passed to the output layer for classification.

\begin{equation}
\label{eq9}
   M_{\text{p}} = X_{\text{p}} \cdot X^T_{\text{p}}  
\end{equation}

\begin{equation}
    N_{\text{p}}(i, j) = \frac{e^{M_{\text{p}}(i,j)}}{\sum_{k=1}^{3} e^{M_{\text{p}}(i,k)}} \text{ for } i, j = 1, 2, 3;
\end{equation}

\begin{equation}
  O_{\text{p}} = N_{\text{p}} \cdot X_{\text{p}}  
\end{equation}

\begin{equation}
   A_{\text{p}} = O_{\text{p}} \odot X_{\text{p}} 
\end{equation}

The attention matrix \( A_{\text{u}} \) is computed separately for \( p = 1^{\text{st}}, 2^{\text{nd}}, \ldots, N^{\text{th}} \) sequences. Finally, for each sequence \( p \), \( A_{\text{p}} \) and \( X_{\text{p}} \) are concatenated and passed to the output layer for classification.

\item \textbf{Multi-Sequence - Self Attention (MS-SA) Framework}: In the MS-SA framework, we apply self-attention to the sequences of each modality separately and use these for classification. In contrast to the MMSS-SA framework, MS-SA utilizes the contextual information of the sequences at the attention level. Let $L$ (text), $V$ (visual), and $L$ (lip region) be the outputs of the dense layers. Three separate attention blocks are required for the three modalities, where each block takes multi-sequence information of a single modality and computes the self-attention matrix. Attention matrices $A_l$, $A_v$, and $A_a$ are computed for lip, visual, and acoustic, respectively. Finally, $A_v$, $A_l$, $A_a$, $V$, $L$, and $A$ are concatenated and passed to the output layer for classification.

\begin{equation}
  M_v = V \cdot V^T  
\end{equation}

\begin{equation}
    N_v(i, j) = \frac{e^{M_v(i,j)}}{\sum_{k=1}^{u} e^{M_v(i,k)}} \text{ for } i, j = 1, \ldots, u
\end{equation}

\begin{equation}
    O_v = N_v \cdot V
\end{equation}

\begin{equation}
    A_v = O_v \odot V
\end{equation}
The attention matrix \( A_p \) is computed for \( p = 1^{\text{st}}, 2^{\text{nd}}, \ldots, u^{\text{th}} \) sequences. Finally, for each sequence \( u \), \( A_p \), and \( X_p \) are concatenated and passed to the output layer with softmax activation for classification.

\end{itemize}

\noindent Further, Figure~\ref{fig:img3} provides further details on the attention computation for our model MMMS-BA discussed in Section $3.2$ of the main paper.

\end{document}